\documentstyle[twoside,fleqn,espcrc2,epsfig]{article}

\title{Free energy of an SU(2) monopole-antimonopole pair
\thanks{Presented by C.~Rebbi. This research was supported
in part under DOE grant DE-FG02-91ER40676 and by the
U.S.~Civilian Research and Development Foundation for
Independent States of FSU (CRDF) award RP1-187.}}

\author{Ch.~Hoelbling$^{\rm a}$,
C.~Rebbi\address{Boston University Physics
    Department\\
    590 Commonwealth Avenue\\
    Boston MA 02215, USA}
        and
V.~A.~Rubakov\address{Institute for Nuclear Research
of the Russian Academy of
  Sciences\\
60th October Anniversary Prospect 7a\\
Moscow 117312, Russian Federation}
}

\begin{document}

\begin{abstract}
We induce an external $\mbox{Z}_2$ monopole-antimonopole pair
in an SU(2) lattice gauge system and measure its free energy as
a way to probe the vacuum structure.  We discuss the motivation
and computational methodology of the investigation and illustrate
our preliminary results.
\end{abstract}

\maketitle

It is well known that some Higgs theories with non-Abelian gauge
group admit stable monopole solutions\cite{tH74,Po74}.  With large gauge
groups, as in grand unified theories, the residual unbroken gauge
group can be non-Abelian.  It is then interesting to determine the
properties of the interaction induced among monopoles, or, as we
will consider in this paper, beteween a monopole and an antimonopole, by the
quantum fluctuations of the unbroken group.  Beyond the relevance that
such interaction may have for the original theory, it can shed
light on the low energy properties of the residual gauge theory
itself.  From the point of view of the unbroken theory, to a very
good approximation the monopoles act as static point sources.
The way to incorporate such sources in an $SU(N)$ lattice gauge theory
was spelled out in Refs.~\cite{UWG80,SS81}, which built on earlier
results established in a seminal paper by 't Hooft~\cite{tH78} and in
Refs.~\cite{MP79,Ya79,MP80}.  We will follow closely the treatment
of Ref.~\cite{SS81}.  In a three dimensional theory, a monopole-antimonopole
pair can be introduced within two cubes of the lattice by the replacement
$ \beta {\rm Tr} U_P \to \beta' {\rm Tr} U_P \equiv z_n \beta {\rm Tr} U_P$
for all the terms in the action corresponding to plaquettes transversed
by a path joining the centers of the two cubes.  $z_n$ stands here for
a non-trivial element of the center of the group $SU(N)$ i.e.~ for one
of the $N^{th}$ roots of unity: $z_n=\exp(2 \pi \imath n /N)$ with
$n=1 \dots N-1$.  In the case of SU(2) which we will study here
the only non-trivial element of the center of the group is -1 and
monopole and antimonopole coincide.  By the redefinition $U \to z_n U$
of an appropriate set of the link variables, the path joining monopole and
antimonopole can be deformed at will.  The path is therefore unphysical and
carries no free energy {\it per se}.  But its end points cannot be moved
without an accompanying change of free energy.  A static monopole-antimonopole
pair is introduced in the four dimensional theory by simply replicating
the above construction in all time slices.  Now the plaquettes with
modified coupling are those transversed by a sheet joining the world
lines of the monopoles.  Again the sheet can be deformed at will.
The insertion of a monopole-antimonopole pair can be reintrepeted
in terms of the electric flux operator introduced in Ref.~\cite{tH78}
and several investigations have been devoted to the study of such
operator in various contexts (see for example Refs.~\cite{BLS81,DGT82}),
but, to the best of our knowledge, no direct calculation of the free
energy of a $Z_2$ monopole-antimonopole pair has ever been attempted.
If the vacuum is characterized by a monopole condensate, one would expect
the behaviour of the free energy as function of separation to exhibit
screening, whereas a $1/r$ behavior or a linearly rising behaviour would
characterize a Coulomb phase or a phase with condensation of electric
charges, respectively.  It is also noteworthy that the monopole and
the antimonopole form two anchors for a center vortex.  Recent investigations
(cfr.~Refs.~\cite{KT98,Co98,Co98a,En98}) have emphasized the role
that such vortices play in confinement.   The calculation which we
present here can be reinterpreted as the calculation of the cost in
free energy to create a center vortex spanning a certain distance within
the lattice.  If such excess free energy quickly saturates (screening),
then the vacuum should indeed exhibit a condensate of center vortices.

The numerical calculation of a free energy is notoriously difficult.
We have been able to obtain reasonably accurate results with acceptable
amounts of CP time by combining a Monte Carlo simulation with
the multihistogram method \cite{FS}.  We consider a modified $SU(2)$ lattice
gauge theory with Wilson action, defined over a $N_x \times N_y
\times N_z \times N_t$ hypercubical lattice with periodic boundary
conditions.  The modification consists in the fact that, for all the
$x-y$ plaquettes $P'$ having a lower vertex with coordinates
$x=0, y=0, 0 < z \le d, 0 \le t < N_t$, the coupling constant
$\beta$ is replaced with $\beta'$.  These are the plaquettes that
cross the sheet joining the worldlines of the monopole and antimonopole
at separation $r=da$ ($a$ being the lattice spacing).
We denote the partition function of this system by $Z(\beta',\beta)$.
We are interested in the free energy
\begin{equation}
F(r)=-\frac{1}{da} \log\Big[\frac{Z(-\beta,\beta)}{Z(\beta,\beta)}\Big]
\end{equation}
Let us define
\begin{equation}
\rho(E)=\int \prod dU \delta[E-\sum_{P'} {\rm Tr} U_{P'}] e^{\sum_{P \ne P'}
\beta {\rm Tr} U_{P} /2}
\end{equation}
If we perform a simulation
with $\beta'$ set to a certain value $\beta_i$ and record in a histogram
the frequency $n_i(E)$ of occurrences of a certain value of $E$, we will
find
\begin{equation}
n_i(E)=  \frac{\rho(E) e^{\beta_i E/2}}{Z(\beta_i,\beta)}
\end{equation}
>From $\rho(E)$ (as yet unknown) we get
\begin{equation}
Z(\beta_i,\beta)=\int dE\, \rho(E) e^{\beta_i E/2}
\end{equation}
The procedure we have followed consists in performing a number $N_h$
of simulations covering the interval $-\beta \le \beta' \le \beta$
with a sufficiently small step to guarantee a good overlap of the
histograms.  We take then
\begin{equation}
\rho(E)=\frac{1}{N_h} \sum_i  n_i(E)  e^{\beta_i E/2} Z_i^{(k)}
\end{equation}
We start from $Z_i^{(0)}=1$ and iterate: the values of $Z_i^{(k)}$
at convergence are proportional to the corresponding $Z(\beta_i,\beta)$.
As a technical improvement in the implementation of the histogram
method, applicable to the case where the measured variable
covers a continous range, we have allocated the values of $E$
to the four neigboring end-points of the histogram intervals
with the weights of a cubic interpolation.  This procedure reduces
substantially the total number of histogram subdivisions one must
use to obtain accurate results.

We illustrate here the results we have obtained with $\beta=2.6$, $N_x=N_y=20$,
$N_z=40$ and the two time extents $N_t=16$ and $N_t=6$, which place the
system in the confined and deconfined phases respectively.  We have
used a combined multi-hit Metropolis overrelaxation algorithm, with
5000 equilibrating iterations and 4000 to 20000 measurements separated by
50 iterations.  The measurements themselves have been performed by
averaging over 384 upgrading steps of the links in the plaquettes $P'$
as a variance reduction technique.  In Figure~1 we show the histograms
for a definite separation of the monopole-antimonopole pair.  In
Figure 2 we show our results for the free energy of the pair.
\begin{figure}
\epsfig{file=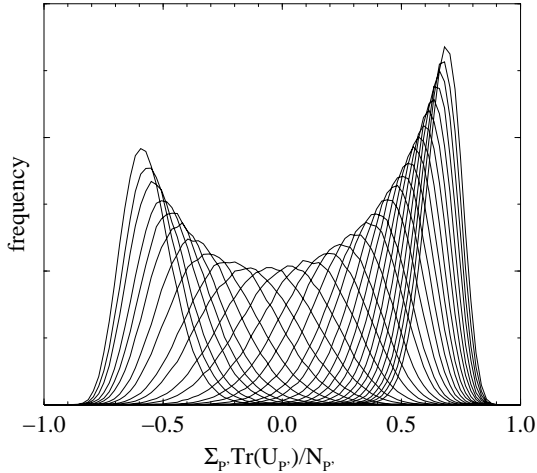,width=7cm}
\caption{The overlap of histograms at $\beta=2.6$ and different
  $\beta^\prime$. The monopole separation is $2a$ and $N_t=6$.}
\end{figure}
\begin{figure}
\epsfig{file=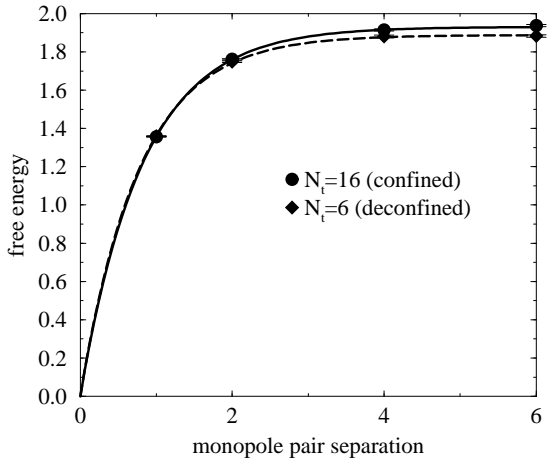,width=7cm}
\caption{Free energy of a monopole-antimonopole pair at different
  separations for $\beta=2.6$ and $N_x\times N_y\times N_z =20\times
  20\times 40$. The circles are with $N_t=16$ in the confined phase, the
  diamonds with $N_t=6$ in the deconfined phase.}
\end{figure}
The calculation is computer intensive, because one must perform separate
calculations for all separations of the pair and for all the intermediate
values of $\beta'$.   However it is quite feasible with present day
computer resources.  We have written a Fortan 90 code which, paying
some attention to the distribution of the data but without resorting
to any special programming tricks like coding critical subroutines
in assembler, runs at approx.~40\% of peak speed on the SGI-Cray Origin
2000, with very satisfactory scaling.  With this performance, the cost
of the data presented in this paper is of the order of a few thousand
processor hours, which is rather modest by today's standards of
supercomputing.

The results in Figure 2 show that the interaction of the pair
is screened both in the confined and deconfined phases.  The lines
in the figure correspond to exponential fits $\exp(-d/l)$ with
$l=0.8300$ and $l=0.7828$ for $N_t=16$ and $N_t=6$, respectively.
We have also computed the free energy of a single monopole
adopting free boundary conditions for the $z=0,N_z$ boundaries of
the lattice (we maintained periodic boundary conditions in all other
directions) and the results are in agreement with the free
energy of the pair for large separation.  Our investigation
is still in progress.  We plan to repeat the calculation with
a smaller value of $\beta$ to verify scaling and to study the behavior
of the free energy of a single monopole as one goes across the
deconfining transition.
It would also be interesting to extend the calculation to other
systems, especially to models which are expected to possess a varied
structure of electric and magnetic confinement phases.

\end{document}